\documentclass[5p,a4paper,sort&compress]{elsarticle}

\usepackage{pifont}
\usepackage{geometry}
\usepackage{fleqn}
\usepackage{multirow}
\usepackage{graphicx}
\usepackage{txfonts}
\usepackage{xcolor}

\newcommand\T{\rule{0pt}{2.6ex}}       
\newcommand\B{\rule[-1.2ex]{0pt}{0pt}} 

\begin{document}

\title{Low-lying single-particle structure of $^{17}$C and the $N= 14$ sub-shell closure}

\author[USC,LPC,UTK,York]{X. Pereira-L\'{o}pez}
\author[USC]{B. Fern\'{a}ndez-Dom\'{i}nguez}
\author[LPC]{F. Delaunay}
\author[LPC]{N.L. Achouri}
\author[LPC]{N.A. Orr}
\author[Surrey]{W.N. Catford}
\author[IPNO]{M. Assi\'{e}}
\author[Bham]{S. Bailey}
\author[GANIL]{B. Bastin}
\author[IPNO]{Y. Blumenfeld}
\author[IFINHH]{R. Borcea}
\author[USC]{M. Caama\~{n}o}
\author[GANIL]{L. Caceres}
\author[GANIL]{E. Cl\'{e}ment}
\author[CEA]{A. Corsi}
\author[Bham]{N. Curtis}
\author[LPC]{Q. Deshayes}
\author[GANIL]{F. Farget}
\author[LNS]{M. Fisichella}
\author[GANIL]{G. de France}
\author[IPNO]{S. Franchoo}
\author[Bham]{M. Freer}
\author[LPC]{J. Gibelin}
\author[CEA]{A. Gillibert}
\author[Regina]{G.F. Grinyer}
\author[IPNO]{F. Hammache}
\author[GANIL]{O. Kamalou}
\author[Surrey]{A. Knapton}
\author[Bham]{T. Kokalova}
\author[CEA]{V. Lapoux}
\author[Sevilla,Sevilla2]{J.A. Lay}
\author[IPNO]{B. Le Crom}
\author[LPC]{S. Leblond}
\author[USC]{J. Lois-Fuentes}
\author[LPC]{F.M. Marqu\'{e}s}
\author[Surrey]{A. Matta}
\author[IPNO]{P. Morfouace}
\author[Sevilla,Sevilla2]{A.M. Moro}
\author[CNS]{T. Otsuka}
\author[GANIL]{J. Pancin}
\author[IPNO]{L. Perrot}
\author[GANIL]{J. Piot}
\author[CEA]{E. Pollacco}
\author[USC]{D. Ramos}
\author[USC,GANIL]{C. Rodr\'{i}guez-Tajes}
\author[GANIL]{T. Roger}
\author[IFINHH]{F. Rotaru}
\author[CEA]{M. S\'{e}noville}
\author[IPNO]{N. de S\'{e}r\'{e}ville}
\author[Bham]{R. Smith}
\author[GANIL]{O. Sorlin}
\author[IFINHH]{M. Stanoiu}
\author[IPNO]{I. Stefan}
\author[GANIL]{C. Stodel}
\author[IPNO]{D. Suzuki}
\author[CNS]{T. Suzuki}
\author[GANIL]{J.C. Thomas}
\author[Surrey]{N. Timofeyuk}
\author[GANIL]{M. Vandebrouck}
\author[Bham]{J. Walshe}
\author[Bham]{C. Wheldon}


\address[USC]{IGFAE and Dpt. de F\'{i}sica de Part\'{i}culas, Univ. of Santiago de Compostela, E-15758, Santiago de Compostela, Spain}
\address[LPC]{LPC Caen, Normandie Universit\'e, ENSICAEN, UNICAEN, CNRS/IN2P3, Caen, France}
\address[UTK]{Department of Physics and Astronomy, University of Tennessee, Knoxville, Tennessee 37996, USA}
\address[York]{Department of Physics, University of York, Heslington, York YO10 5DD, United Kingdom}
\address[Surrey]{Department of Physics, University of Surrey, Guildford GU2 5XH, UK}
\address[IPNO]{Universit\'{e} Paris-Saclay, CNRS/IN2P3, IJCLab, 91405 Orsay, France}
\address[Bham]{School of Physics and Astronomy, University of Birmingham, Birmingham B15 2TT, UK}
\address[GANIL]{GANIL, CEA/DRF-CNRS/IN2P3, Bd. Henri Becquerel, BP 55027, F-14076 Caen, France}
\address[CEA]{D\'{e}partement de Physique Nucl\'{e}aire, IRFU, CEA, Universit\'{e} Paris-Saclay, F-91191 Gif-sur-Yvette, France}
\address[LNS]{INFN, Laboratori Nazionali del Sud, Via S. Sofia 44, Catania, Italy}
\address[Regina]{Department of Physics, University of Regina, Regina, SK S4S 0A2, Canada}
\address[Sevilla]{Departamento de FAMN, Facultad de F\'{i}sica, Universidad de Sevilla, Apdo. 1065, E-41080 Sevilla, Spain}
\address[Sevilla2]{Instituto Interuniversitario Carlos I de F\'{i}sica Te\'{o}rica y Computacional \(iC1\), Apdo. 1065, E-41080 Sevilla, Spain}
\address[CNS]{CNS, University of Tokyo, 7-3-1 Hongo, Bunkyo-ku, Tokyo, Japan}
\address[IFINHH]{IFIN-HH, P. O. Box MG-6, 76900 Bucharest-Magurele, Romania}

\date{\today}

\begin{abstract}

The first investigation of the single-particle structure of the bound states of $^{17}$C , via the $d(^{16}$C$,p)$ transfer reaction, has been undertaken. The measured angular distributions confirm the spin-parity assignments of $1/2^+$ and $5/2^+$ for the excited states located at 217 and 335~keV, respectively. The spectroscopic factors deduced for these states exhibit a marked single-particle character, in agreement with shell model and particle-core model calculations, and combined with their near degeneracy in energy provide clear evidence for the absence of the $N=14$ sub-shell closure. The very small spectroscopic factor found for the $3/2^+$ ground state is consistent with theoretical predictions and indicates that the $\nu 1d_{3/2}$ strength is carried by unbound states.
With a dominant $\ell = 0$ valence neutron configuration and a very low separation energy, the $1/2^+$ excited state is a one-neutron halo candidate.
\end{abstract}



\maketitle

The evolution of nuclear shell structure with isospin has been the focus of much attention over the last two decades (see, for example, Ref.~\cite{OtsukaRMP}).  Of particular interest are the light neutron-rich nuclei where, experimentally, radioactive beams of sufficient intensity have become available to allow systems to be studied along isotopic chains to the limits of nuclear binding.  In parallel, increasingly sophisticated shell model interactions \cite{TSuzukiI,TSuzukiII,VMU,YSOX,CCEI} have been developed in order to understand the properties of these nuclei.

Studies of the neutron-rich oxygen isotopes have demonstrated the emergence of $N=14$ and 16 as clear sub-shell closures, with $^{22}$O and $^{24}$O considered to be doubly magic \cite{Ozawa,MStanoiu,Becheva,ZElekes-O23,Hoffman-N16,Hoffman-O24,BFD-O21,TshooI,TshooII}.  Intriguingly, a comparison of the systematics of the energies of the 2$^{+}$ levels in the oxygen and carbon isotopic chains suggest that the $N=14$ gap is no longer present for $Z=6$ \cite{MStanoiu-N=14Cchain}.  Schematic mechanisms have been proposed to explain this in terms of the proton-neutron interactions and the absence of $1p_{1/2}$ protons in carbon \cite{OtsukaRMP,MStanoiu-N=14Cchain}.  Shell model calculations employing phenomenological two-body interactions have struggled to explain the structure of the neutron-rich O and C isotopes simultaneously.  Indeed, ad hoc $Z$-dependent reductions in the monopole terms of these interactions have been invoked in order to reproduce experiment \cite{MStanoiu,DSohler,MStanoiu-N=14Cchain}.  

More realistic Hamiltonians, such as the SFO-tls \cite{TSuzukiII} and YSOX \cite{YSOX} interactions which include an improved treatment of the tensor interaction, have been developed to describe $psd$ shell nuclei.
These interactions are able, for example, to reproduce the exotic character of the magnetic dipole transitions in $^{17}$C \cite{TSuzukiII} and to describe the drip lines for $Z=6$ and 8 \cite{YSOX}.
In a more fundamental approach applied to this region, effective interactions have been derived from chiral two- and three-nucleon forces using the \textit{ab-initio} coupled-cluster method \cite{CCEI}. Without the tuning of any parameters level schemes of neutron-rich C and O isotopes comparable to those obtained with phenomenological effective interactions have been obtained. The No-Core Shell Model approach has also begun to be employed in this region -- including the neutron-rich C isotopes \cite{PetriLifetimeC16,SmalleyLifetimeC17,VossLifetimeC18} -- whereby the explicit inclusion of three-nucleon forces and, for weakly bound systems, the continuum is required. 

In order to provide for a more detailed understanding of the shell-structure of the neutron-rich carbon isotopes and to test shell-model interactions, information on the single-particle structure of these nuclei is of key interest.   In the present work attention is concentrated on $^{17}$C, where the structure of the low-lying levels should be governed by the valence neutrons occupying the $1d_{5/2}$, $2s_{1/2}$ and $1d_{3/2}$ single-particle orbitals. 

Experimentally, the structure of $^{17}$C has been the object of a number of investigations \cite{MStanoiu-N=14Cchain,Ogawa,Maddalena,Kondo,ZElekes-LowlyingC1719,Bohlen,DSuzuki-LifetimeC17,SmalleyLifetimeC17,Ueno,SauvanPLB,SauvanPRC,UDatta}. The ground state has a well established spin-parity assignment of $3/2^{+}$ \cite{Ogawa, SauvanPLB,SauvanPRC,Maddalena} and has been seen via neutron removal \cite{Maddalena} to be built from several configurations: a dominant $^{16}\mbox{C}(2^+_1) \otimes \nu 1d_{5/2}$ component, a smaller $^{16}\mbox{C}(2^+_1) \otimes \nu 2s_{1/2}$ component, a $^{16}\mbox{C}(0^+_{g.s.}) \otimes \nu 1d_{3/2}$ configuration and $\ell=2$ components coupled to the $2^+_2$, $3^+_1$ and $4^+_1$ levels. The measured partial cross section for neutron removal leading to the ground state of $^{16}$C is, however, an order of magnitude higher than expected from shell model calculations \cite{Maddalena}. This result is also in clear disagreement with that obtained from the Coulomb dissociation  \cite{UDatta} of $^{17}$C which exhibits a small cross section to the ground state of $^{16}$C. Significantly, if correct, a large cross section for the neutron removal would suggest that the $N=16$ gap is significantly diminished.

In terms of the low-lying level structure of $^{17}$C, bound excited states -- $S_n=734\pm18$~keV \cite{AME16} -- have been established to lie at 0.217(1) and 0.332(2) MeV \cite{ENSDF} and have been observed in a variety of reaction studies \cite{MStanoiu-N=14Cchain,ZElekes-LowlyingC1719,Bohlen,DSuzuki-LifetimeC17,SmalleyLifetimeC17} as well as $\beta$-decay \cite{Ueno}.  Momentum distributions measured in single-neutron removal from $^{18}$C were found to be consistent with spin-parity assignments of 1/2$^{+}$ and 5/2$^{+}$ respectively \cite{Kondo}.  In the case of the former state, the very small binding energy -- $S_n=517\pm18$~keV -- and $2s_{1/2}$ neutron configuration suggest that it may exhibit a halo, as hinted at by lifetime measurements \cite{DSuzuki-LifetimeC17,SmalleyLifetimeC17}.
In the present work, the investigation of the single-particle structure of the bound states of $^{17}$C using the $(d,p)$ single-neutron transfer reaction in inverse kinematics with an energetic secondary $^{16}$C beam is reported.

\begin{figure}
\includegraphics[scale=0.45]{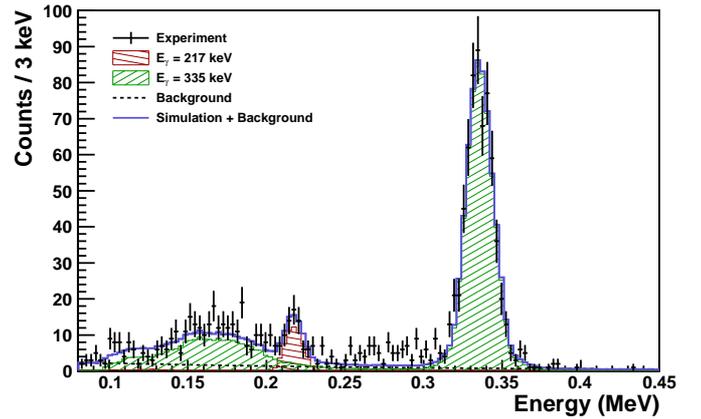}%
\caption{\label{gammas} Doppler-corrected $\gamma$-ray energy spectrum in coincidence with protons leading to bound states in $^{17}$C compared to the simulated response for 217 and 335~keV $\gamma$-rays and a background component.}
\end{figure}

The $^{16}$C beam was produced using an intense 50 AMeV $^{18}$O ($\sim$0.8 p$\mu$A) primary beam, delivered by the GANIL coupled-cyclotron facility, incident on a Be production target (2~mm thick) tilted at 37$^{\circ}$.  The LISE3 spectrometer, equipped with a thick Be achromatic degrader (1.9~mm), was employed to select, purify and slow down the secondary beam. The $^{16}$C beam so obtained was 100\% pure, with an intensity of $\sim5 \times 10^{4}$ pps. The mean energy of the beam was 17.2~MeV/nucleon and its energy spread, as defined by the spectrometer acceptance, was 2.5\%.

Owing to the relatively limited optical characteristics of the secondary beam, the beam particles were tracked event by event onto the secondary CD$_{2}$ target  using a set of two multiwire proportional chambers \cite{CATS}. These detectors also provided a beam time reference as well as the means to determine the number of incoming ions. The CD$_{2}$ target was surrounded by the TIARA Silicon array \cite{TIARA} comprising an octagonal double-layered barrel of resistive strip detectors spanning 36$^\circ$ to 144$^\circ$ and an annular double-sided silicon strip detector covering the most backward angles (144$^\circ$ - 169$^\circ$). Four germanium clover detectors of the EXOGAM array \cite{EXOGAM} were placed at 90$^\circ$ surrounding the barrel in a compact arrangement 55 mm from the centre of the target.  The front faces of the detectors spanned angles from 45$^{\circ}$ to 135$^{\circ}$. The photopeak efficiency, including the Lorentz boost, was determined to be 14.8$\pm$0.2 and 13.7$\pm$0.2\% at 217 and 335 keV, respectively. For the former the efficiency takes into account the effects resulting from the lifetime of the state ($\tau$ = 528$^{+21}_{-14}$~ps \cite{SmalleyLifetimeC17}). Source measurements coupled with GEANT4~\cite{GEANT4} based simulations determined that the lifetime resulted in a reduction of the detection efficiency from 16.5 to 14.8\% without any appreciable lineshape asymetry or increase in the observable width of the 217~keV $\gamma$-ray line.  Atomic-number identification of the non-interacting beam particles and beam-like residues was achieved by measuring energy loss, residual energy and time of flight using a Si-Si-CsI telescope located at zero degrees 33 cm downstream of the target \cite{CHARISSA}.

The protons from the $(d,p)$ reaction to the three bound states of $^{17}$C had energies within a range of some 100~keV and could not be separated given the energy resolution of TIARA. Therefore, protons from transfer to each excited state were selected by gating on the corresponding $\gamma$-ray line \cite{GWilson}. The Doppler corrected $\gamma$-ray energy spectrum measured in coincidence with protons populating the bound states (selected using the kinematic locus in energy versus angle \cite{XPereiraPhD}) and registered in coincidence with $Z=6$ beam-like residues at zero degrees is shown in Figure \ref{gammas}. We note, as discussed below, that shell model calculations do not predict any significant population of unbound states below 2.5~MeV in $^{17}$C.  Figure \ref{gammas} also displays the lineshapes, derived from GEANT4~\cite{GEANT4} based simulations, for the detection of 217-keV and 335-keV $\gamma$-rays. Note that each of the bound excited states decay exclusively by a direct transition to the ground state \cite{ENSDF}.

\begin{figure}
\includegraphics[scale=0.63]{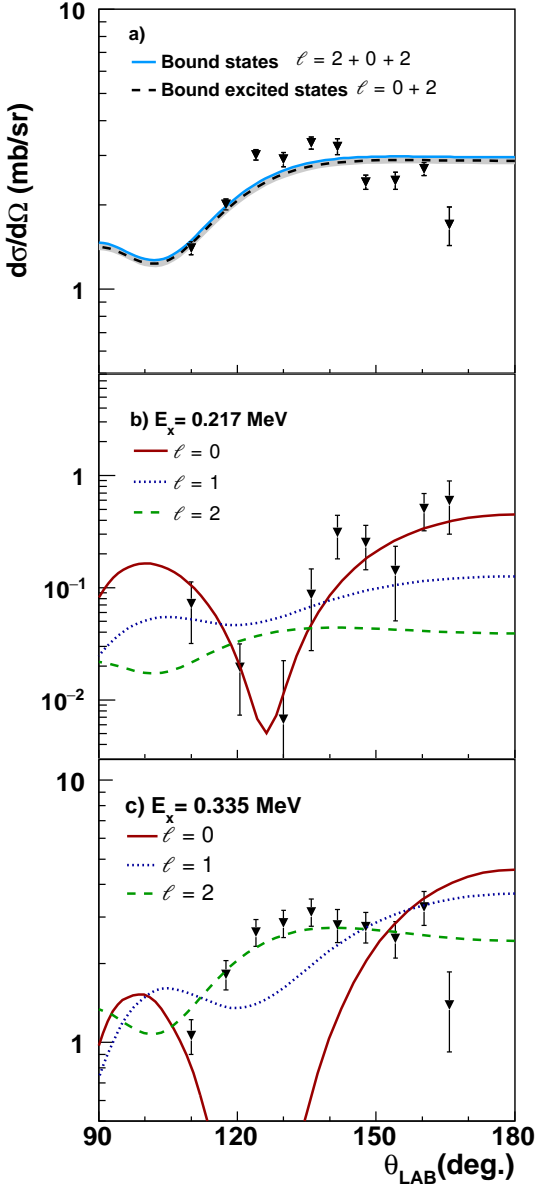}%
\caption{\label{exdistros} Experimental angular distributions in the laboratory frame for protons leading to (a) the three bound states and (b) to the first and (c) second excited states of $^{17}$C, compared with ADWA calculations assuming angular momentum transfers of $\ell=0$, 1 and 2. Note that for the first excited state (b), owing to the limited statistics in the region between 120-135 degrees a broader angular binning was employed.  In panel (a) the dashed line is the ADWA angular distribution for the sum of the first and second excited states while the shaded band represents the associated statistical uncertainty, excluding the systematic uncertainty from reaction modelling (see text).  The solid line in (a) represents the additional contribution for the $\ell=2$ transfer to the ground state.} 
\end{figure}

To construct differential angular cross sections (d$\sigma$/d$\Omega$) for the transfer to the excited states, the $\gamma$-ray energy spectrum in coincidence with protons detected in a given angular bin was fitted by the sum of the response functions for the two $\gamma$-ray lines and background. The integral of the fitted response function, corrected by the $\gamma$-ray detection efficiency, gave the number of protons for the population of the corresponding state.
The absolute cross section was derived using the deuterium thickness of the target\footnote{The elastic scattering allowed the $^{1}$H contamination of the target to be determined to be less than 1\% \cite{XPereiraPhD}.} and the measured number of $^{16}$C ions. The target thickness (1.37$\pm$0.04~mg/cm$^2$) was determined via energy loss measurements using an alpha source. In addition, the $(d,d)$ elastic scattering angular distribution was employed \cite{XPereiraPhD}  to obtain an independent determination of the normalization \cite{GWilson}, which was found to be consistent with that obtained from the target thickness and the number of incident ions. We note that no corrections were applied to account for proton-$\gamma$-ray angular correlations \cite{WNC-Euroschool} as the effect on the proton distributions was found to be similar to the statistical uncertainties owing to the very broad angular coverage of the Ge detector array.

The angular distributions derived for the two excited states are presented in Figure \ref{exdistros}. In order to determine the transferred angular momentum $\ell$ and deduce spectroscopic factors they are compared to results from Adiabatic Distorted Wave Approximation (ADWA) \cite{ADWA} calculations, which take into account the effect of deuteron breakup.  The Koning-Delaroche (KD) \cite{KoningDelaroche} and Chapel-Hill (CH89) \cite{CH89} parameterizations were employed for the $p$+$^{16}$C and $n$+$^{16}$C optical potentials used to construct the $d$+$^{16}$C adiabatic potential and for the $p$+$^{17}$C potential in the exit channel. We note that both nucleon-nucleus potential parameterizations described extremely well the $p$+$^{16}$C elastic scattering angular distribution measured here with a CH$_2$ target \cite{XPereiraPhD}.

The adiabatic potential was built within the Johnson-Tandy finite range prescription \cite{JohnsonTandy} using a deuteron wave-function obtained with the Reid soft-core $np$ interaction.
The ADWA calculations were performed using the TWOFNR code \cite{TWOFNR} with the zero-range approximation with a standard finite-range correction and a standard non-locality correction in the $p$+$^{17}$C exit channel.
The $\langle ^{17}\mbox{C} | ^{16}\mbox{C}\rangle$ overlaps were derived from neutron wave functions in a Woods-Saxon potential with standard radius and diffuseness parameters ($r_0=1.25$ fm and $a=0.65$ fm) and the depth adjusted to reproduce the experimental neutron separation energy.
The results displayed in Figure \ref{exdistros} correspond to the KD potential.  Those obtained using the CH89 parameterization provided for angular distributions with very similar shapes.

The angular distribution for the first excited state, including most notably the characteristic pronounced minimum at around 125$^\circ$ in the laboratory frame, corresponds clearly to a $\ell=0$ angular momentum transfer, leaving 1/2$^+$ as the only possible spin-parity assignment, in agreement with that suggested by Ref. \cite{Kondo}.
The spectroscopic factor, obtained by a $\chi^2$ minimisation of the normalisation, is 0.64$\pm$0.18 and 0.80$\pm$0.22 for the KD and CH89 parameterizations respectively.  

The angular distribution for the second excited state provides for an obvious $\ell=2$ assignment. On the basis of comparison with the structure calculations presented below, a spin-parity assignment of 5/2$^+$ is clearly favoured, again in agreement with the previous suggested assignment \cite{Kondo}.
The corresponding spectroscopic factor was deduced to be 0.62$\pm$0.13 for both the KD and CH89 parameterizations.

The angular distribution for the sum of all three bound states was obtained by fitting the reconstructed excitation energy spectrum for each angular bin without requiring any coincident $\gamma$-rays.  The fit took into account the excitation energy resolution as well as the tails of any resonances and three-body phase space arising from deuteron breakup.  The result is displayed in Figure \ref{exdistros}a). 
Given the established $J^\pi$=3/2$^+$ assignment for the ground state of $^{17}$C, the angular momentum transfer should be $\ell=2$.
The extraction of the corresponding spectroscopic factor is not, however, straightforward
as it is expected to be rather small (see shell model predictions below).  More specifically, the very large $^{16}\mbox{C}(2^+)\otimes \nu 1d_{5/2}$ component is not acessible via single-step neutron transfer onto the $^{16}\mbox{C}$ ground state\footnote{Coupled-channels calculations allowing two-step processes via the $^{16}$C(2$^+$) state suggest that their effect is less than the uncertainty on the ground state spectroscopic factor.}. 
Indeed, as may be seen in Figure \ref{exdistros}a), the ADWA angular distribution for the sum of the two bound excited states, computed using the spectroscopic factors deduced above, describes well the inclusive angular distribution (i.e., for the three bound states).  A slightly improved description is obtained when including a small additional component arising from the $\ell=2$ transfer to the ground state which corresponds to a spectroscopic factor of 0.03$^{+0.05}_{-0.03}$ for both the KD and CH89 potentials.

\begin{figure*}
  \begin{center}
    \includegraphics[scale=0.72]{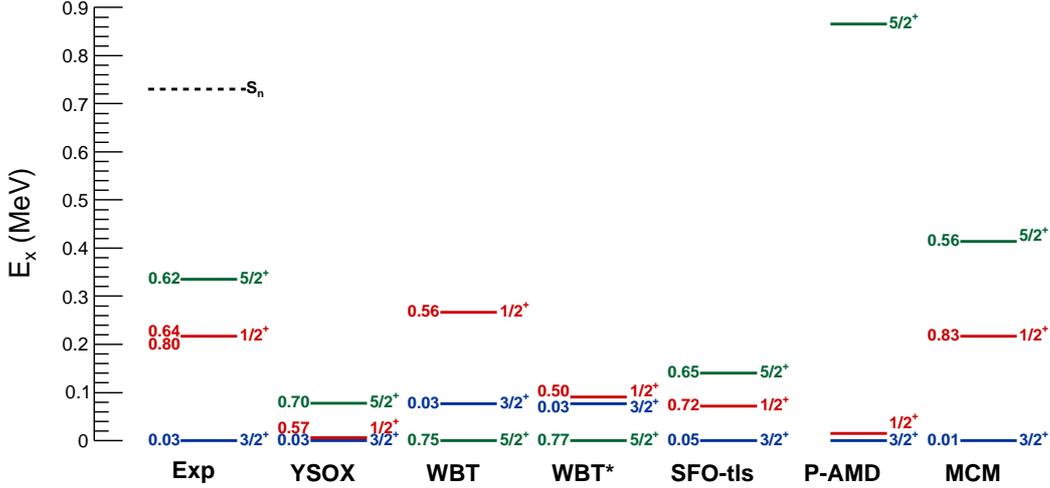}%
\caption{\label{nrj} The $^{17}$C level scheme and spectroscopic factors from the present study (uncertainties are listed in Table \ref{results}) compared to shell model calculations using the WBT \cite{WarburtonBrown}, modified WBT (WBT*) \cite{MStanoiu-N=14Cchain}, YSOX \cite{YSOX} and SFO-tls \cite{TSuzukiII} interactions, and to the P-AMD \cite{PAMD} and MCM \cite{MCM17C} particle-core models (see text).  In the case of the 1/2$^+$ level, the experimental spectrocopic factors for the KD and CH89 potentials are shown.}
\end{center}
\end{figure*}

The measured excitation energies, spin-parity assignments and $^{16}\mbox{C}(0^+_{g.s.}) \otimes \nu nlj$ spectroscopic factors are listed in Table \ref{results} and included in Figure \ref{nrj}. The uncertainties ascribed to the spectroscopic factors include the uncertainty arising from the reaction modelling, which is estimated to be 20\% \cite{LeeSF}, statistical fitting errors and the uncertainty in the target thickness (3 \%).

\begin{table*}
\begin{center}
  \caption{\label{results} Spin-parity assignments $J^\pi$, excitation energies $E_{x}$ (keV) and spectroscopic factors $C^2S$ for the corresponding $\ell$ transfers for the bound states of $^{17}$C using different optical model potentials, compared to shell model predictions for the YSOX, WBT, WBT* and SFO-tls interactions, and to the MCM and P-AMD particle-core models (see text). The quoted uncertainties include statistical fitting errors, systematic uncertainties from the reaction modelling (20 \% \cite{LeeSF}) and the error on the target thickness (3 \%).}
\begin{tabular}{ccc cc cc cc cc cc cc c}
  \hline
  \hline
  & &  & KD & CH89 & \multicolumn{2}{c}{YSOX} & \multicolumn{2}{c}{WBT} & \multicolumn{2}{c}{WBT*} & \multicolumn{2}{c}{SFO-tls} & \multicolumn{2}{c}{MCM} & P-AMD \T\B \\
  \hline
  $J^{\pi}$ & $E_{x}^{exp}$ & $\ell$ &$C^2S^{exp}$ & $C^2S^{exp}$ & $E_{x}$ & $C^2S$ & $E_{x}$ & $C^2S$ & $E_{x}$ & $C^2S$ & $E_{x}$ & $C^2S$ & $E_{x}$ & $C^2S$ &  $E_{x}$ \T\B\\
		\hline
		$3/2^{+}$ & 0 & 2 & 0.03$^{+0.05}_{-0.03}$ & 0.03$^{+0.05}_{-0.03}$  & 0 & 0.03 & 77 & 0.03 & 77 & 0.03 & 0 & 0.05 & 0 & 0.01 & 0 \T \\
		$1/2^{+}$ & 217(1) & 0 & 0.64$\pm$0.18 & 0.80$\pm$0.22 & 6 & 0.57 & 267 & 0.56 & 91 & 0.50 & 72 & 0.72 & 207 & 0.83 & 15 \\
		$5/2^{+}$ & 335(1) & 2 & 0.62$\pm$0.13 & 0.62$\pm$0.13 & 78 & 0.70 & 0 & 0.75 & 0 & 0.77 & 140 & 0.65 & 414 & 0.56 & 866 \B\\
		\hline
                \hline
\end{tabular}
\end{center}
\end{table*}


The results are compared in the following to shell model calculations performed with the WBT \cite{WarburtonBrown} and WBT* \cite{MStanoiu-N=14Cchain} (with the neutron-neutron two-body matrix elements scaled to reproduce experimental data) interactions in the $spsdpf$ model space and with the YSOX \cite{YSOX} and SFO-tls \cite{TSuzukiII} interactions in the $psd$ model space (Table \ref{results} and Figure \ref{nrj}).
In addition, comparison is made with two particle-core models: the Microscopic Cluster Model (MCM) \cite{MCM17C} and the semi-microscopic Particle Antisymmetrized Molecular Dynamics model (P-AMD) \cite{PAMD}. In the MCM model, the strength of the Majorana part of the nucleon-nucleon interaction in odd partial waves and the spin-orbit amplitude have been adjusted to reproduce the energies of the ground and first excited state. The P-AMD is essentially a particle-rotor model in which the $^{16}$C core-particle potentials are obtained by folding an effective nucleon-nucleon interaction with the core densities obtained from Antisymmetrized Molecular Dynamics calculations. As such, the core-particle potentials have no free parameters. Both the MCM and P-AMD models include excited states of the core. The P-AMD calculation performed here for $^{17}$C is similar to those undertaken previously for $^{11}$Be and $^{19}$C \cite{PAMD}. The results are listed in Table \ref {results} and Figure \ref{nrj}.
Contrary to the MCM, the P-AMD model does not include antisymmetrization between the valence neutron and those of the core. Consequently spectroscopic factors cannot be directly extracted from the latter and are not presented here.

All of the shell model calculations, as well as the MCM model, predict low-lying 1/2$^{+}$, 3/2$^{+}$, and 5/2$^{+}$ levels.  Whilst the ordering of states is not always reproduced, this is not surprising given the typical deviations observed for nuclei in this region ($\sim$300~keV \cite{WarburtonBrown}).  More significantly, the spectroscopic factors are in good accord with the experimentally deduced values

As discussed earlier, the present study finds that the $^{17}\mbox{C}$ ground state exhibits a very small $^{16}\mbox{C}(0^+_{g.s.}) \otimes \nu 1d_{3/2}$ spectroscopic factor. In addition to being in line with theory, this result is consistent with the small Coulomb dissociation cross section of $^{17}$C to the ground state of $^{16}$C \cite{UDatta} and one of the intermediate energy neutron removal studies \cite{SauvanPLB,SauvanPRC}.
Importantly, the small spectroscopic factor deduced here indicates that essentially all the $\nu 1d_{3/2}$ strength, which carries information on the $N=16$ sub-shell closure, is carried by unbound states.  This is supported by theory whereby all the shell model calculations predict that the strength lies in states above 2~MeV.


The spectroscopic factors deduced here for the two excited states exhaust a large fraction of the available single-particle strength.  
More specifically, taking the spectroscopic factor sum rule combining nucleon addition and nucleon removal on the same nucleus \cite{Satchler} and given the spectroscopic factors measured for neutron removal from $^{16}$C \cite{Maddalena}, the spectroscopic strength available for $^{16}$C$(d,p)$ populating the $\nu 1d_{5/2}$ and $\nu 2s_{1/2}$ orbitals is 0.79 and 0.72, respectively.
The spectroscopic factors deduced here therefore exhaust $\sim$80\% of the available $1d_{5/2}$ strength and $\sim$100\% of the available $2s_{1/2}$ strength.  This indicates that no unbound 5/2$^+$ and 1/2$^+$ states with significant single-particle strength should be expected in $^{17}$C.

The single-particle character of the almost degenerate 1/2$^+$ and 5/2$^+$ levels strongly suggests that $N=14$ shell closure does not occur in $^{17}$C as the WBT interaction, for example, predicts (Figure 5 of Ref.~\cite{MStanoiu-N=14Cchain}).  In the case of the SFO-tls interaction, which also reproduces well the low-lying levels of $^{17}$C, the effective single-particle energies (ESPE) are $\epsilon_{\nu 2s1/2} = -1.20$ MeV and $\epsilon_{\nu 1d5/2}=-1.84$~MeV, confirming the near degeneracy of the two orbitals.
A rigorous procedure to extract ESPE from measurements combines nucleon removal and addition \cite{Baranger,Signoracci}. This procedure, initially introduced for a closed-shell nucleus, is in fact valid for any $0^+$ nucleus \cite{OtsukaRMP}.
For $^{16}$C, combining the present results for neutron addition with those for neutron removal \cite{Maddalena}, ESPE of $\epsilon_{\nu 2s1/2} = -1.56\pm0.18$ and $\epsilon_{\nu 1d5/2}=-1.58\pm0.17$~MeV are obtained\footnote{In the case of the $^{17}$C 1/2$^+$ state, where the KD and CH89 spectroscopic factors differ, the weighted average was used (as is also the case for the calculation of the ANC).}, demonstrating that the $N=14$ sub-shell closure is also absent in $^{16}$C.
In the case of the SFO-tls interaction, $\epsilon_{\nu 2s1/2} = -1.35$ and $\epsilon_{\nu 1d5/2}=-1.34$~MeV are predicted for $^{16}$C.

In terms of the particle-core models, the MCM predicts spectroscopic factors in agreement with the experiment. With the parameters adjusted to reproduce the energies of the ground and $1/2^+$ states, the energy of the $5/2^+$ state agrees well with the experiment.
In the P-AMD model, the excitation energy of the $1/2^+$ state is somewhat underestimated while that of the $5/2^+$ state is too high. However, as pointed out above, this model has no free parameters in the core-neutron potentials.


The MCM model predicts the three bound states to have significant excited core components \cite{MCM17C}.
In particular, it indicates, as expected (see above), that the $3/2^+$ ground state is dominated by the $^{16}$C$(2^+_1) \otimes \nu 1d_{5/2}$ configuration ($C^2S\sim$1.3) and contains a significant $4^+_1$ core excitation ($C^2S\sim$0.4). The $5/2^+$ state is predicted to have a large $4^+_1$ admixture ($C^2S\sim$0.95), while the main core excitation in the $1/2^+$ state involves the $2^+_2$ state ($C^2S\sim$0.4). Interestingly, the P-AMD model, which overestimates the energy of the $5/2^+$ state, includes only the coupling to the $2^+_1$ state of $^{16}$C.
The importance of including the $2^+_1$ and $4^+_1$ states of the core for the correct description of the level scheme was also pointed out more generally by multichannel algebraic scattering calculations \cite{MCAS}.

\begin{figure}
\includegraphics[scale=0.45]{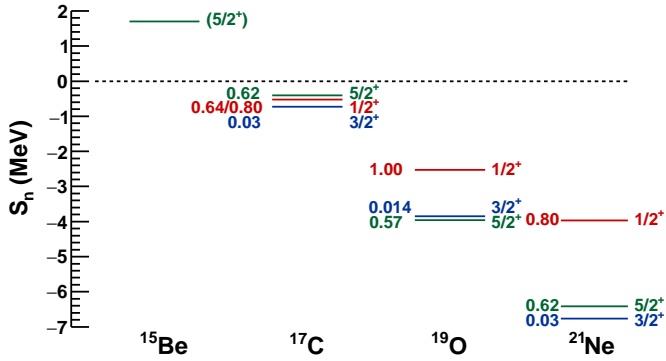}%
\caption{\label{N11} Neutron binding energies of the positive parity states in the $N=11$ isotonic chain and the corresponding experimental spectroscopic factors where available: $^{15}$Be \cite{Be15}, $^{17}$C (this work - Table~\ref{results}), $^{19}$O \cite{ELevelsA19} and $^{21}$Ne \cite{ELevelsA21_44}.}
\end{figure}

Figure \ref{N11} displays the neutron separation energies of the lowest $1/2^+$, $3/2^+$ and $5/2^+$ states in the even-$Z$ neutron-rich $N=11$ isotones. Where available, the core$(0^+_{g.s.}) \otimes \nu nlj$ spectroscopic factors derived from $(d,p)$ reaction studies are also indicated.
In unbound $^{15}$Be only one state has been observed and its spin-parity tentatively assigned to be $5/2^+$ \cite{Be15}.
The three other $N=11$ isotones exhibit striking similarities. In particular, the lowest $3/2^+$ and $5/2^+$ states are almost degenerate. In addition, the $3/2^+$ states exhibit very small spectroscopic factors ($\leq 0.05$), while the $5/2^+$ and $1/2^{+}$ states carry considerable single-particle strength ($\sim0.6 - 1.0$). 
The energy of the $1/2^+$ state with respect to the $5/2^+$ state decreases from around 2.4~MeV in $^{21}$Ne to  $-0.12$ MeV in $^{17}$C. Based on these systematics, one would expect the ground state of $^{15}$Be to have a spin-parity of $1/2^+$.

It has been shown by Hoffman \textit{et al.} \cite{HoffmanKaySchiffer} and Hamamoto \cite{Hamamoto} that the effects of finite binding play an important role in determining the ordering of levels in loosely bound light nuclei. In particular, the decrease here of the energy difference between the $5/2^+$ and $1/2^+$ states from $^{21}$Ne to $^{17}$C might, to a large extent, be due to the different behaviour of the $2s_{1/2}$ and $1d_{5/2}$ neutron orbitals as their binding energy decreases. This can be investigated, in the spirit of Ref.~\cite{HoffmanKaySchiffer}, by simple calculations in a Woods-Saxon potential. Starting with a potential geometry identical to that used in Ref.~\cite{HoffmanKaySchiffer} ($r_{0}=1.25$ fm, $a=0.63$ fm),  the depths of the central and spin-orbit terms have been adjusted such that the energies of the $1/2^+$ and $5/2^+$ states of $^{19}$O were reproduced. For $^{17}$C and $^{21}$Ne the strength of the central term was varied so as to reproduce the energy of the $5/2^+$ state and the energy of the $1/2^+$ level was computed. The latter was found to lie within $\sim$20 and $\sim$200~keV of experiment for $^{17}$C and $^{21}$Ne, respectively -- the deviation in both cases being small compared to the actual separation energy.

With a dominant $\ell=0$ valence neutron configuration, as evidenced by the large $\nu 2s_{1/2}$ spectroscopic strength observed here, and a low separation energy ($S_n=517\pm$18~keV), the 1/2$^+$ state is a good candidate for a neutron halo.  One of the key quantities in defining a neutron halo is the probability $P_n(r>\rho)$ of finding the valence neutron at a radius ($r$) larger than the classical turning point, $\rho$ \cite{RiisagerPhysScrHalos}.
For a given state, $P_n(r>\rho)$ can be evaluated from the density $r^2 R^2$, where $R$ is the radial part of the core-neutron relative wave function.
Here we have used the P-AMD model which correctly accounts for the asymptotic behaviour of wave functions.
It may be noted that the P-AMD overestimates the separation energy of the $1/2^+$ state and underestimates that of the $5/2^+$ state.  As such, the microscopic core-neutron potentials were rescaled to obtain the two states at the experimental energies, which did not lead to significant changes in the content of the wave functions.

\begin{figure}
\includegraphics[scale=0.45]{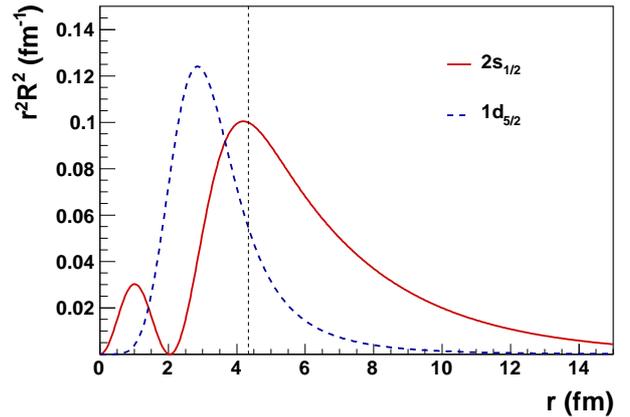}
\caption{\label{FigHalo} Density $r^2 R^2$ of the $2s_{1/2}$ valence neutron of the $1/2^+$ state with $S_n=518$~keV (solid line) and of the $1d_{5/2}$ valence neutron of the $5/2^+$ state with $S_n=400$ keV (dashed line), calculated with the P-AMD model including excitations of the $^{16}$C core ($R$ being the radial part of the wave function). The vertical dashed line indicates the classical turning point $\rho = 4.3$ fm estimated for the $^{16}$C-neutron potential.}
\end{figure}

The densities of the valence neutron in the $^{16}$C$(0^+_{gs})\otimes2s_{1/2}$ component of the 1/2$^+$ state and in the $^{16}$C$(0^+_{gs})\otimes1d_{5/2}$ component of the 5/2$^+$ state from the P-AMD model with core excitations are shown in Figure~\ref{FigHalo}. These components are the ones directly probed in deuteron stripping.
Using a $^{16}$C rms radius of 2.7~fm \cite{OzawaInteractionXS}, a classical turning point, $\rho$, of 4.3~fm is estimated \cite{RiisagerPhysScrHalos} for $^{16}$C$+n$, as indicated in Figure~\ref{FigHalo}.

Although it is somewhat more bound, the density distribution of the $\ell=0$ neutron of the 1/2$^+$ state extends much further out than that for the $\ell=2$ neutron of the 5/2$^+$ state.  Quantitatively, for the 1/2$^+$ state, including core excitation components, $P_n(r>\rho) = 45$\% and the valence neutron rms radius is 5.8~fm, whereas for the $5/2^+$ state (again including core excited components) $P_n(r > \rho) = 37$\% and the valence neutron rms radius is 4.5~fm. 


A direct measure of the normalization of the valence neutron wave-function outside of the $^{16}$C core can be provided by the Asymptotic Normalization Coefficient (ANC) \cite{ANC}. The ANC of a given state is defined as the product of the square root of the experimental spectroscopic factor and the single-particle ANC of the normalized bound-state wave function used in the reaction calculation, provided that the reaction is peripheral.
In addition to providing the normalization of the wave functions outside of the core-particle potential, neutron ANCs can be used to predict the widths of resonances in the unbound proton-rich mirror nucleus \cite{BFD-25P}, which for $^{17}$C is the potential three-proton emitter $^{17}$Na \cite{MCM17C}.  The peripherality of the present reaction was checked by studying the dependence of the ANC on the geometric parameters of the neutron binding potential. For changes of $\pm$10\% in the radius ($r_0$) and diffuseness ($a$) parameters, leading to variations in the spectroscopic factors of up to 15 \%, the ANC changes only by $\sim$1\%. The absolute ANC values derived here for the ground, first and second excited states of $^{17}$C were 0.02$\pm$0.01, 0.78$\pm$0.08 and 0.048$\pm$0.04 fm$^{-1/2}$, respectively. 

In the asymptotic region ($r \gtrsim 15$ fm), the ANCs extracted here imply a factor $\sim$20 higher density for the $1/2^+$ state compared to that of the $5/2^+$ state.
The experimentally determined ANC values can be directly compared to theoretical calculations.
In the P-AMD model, the absolute value of the ANC of the $1/2^+$ state is 0.76 fm$^{-1/2}$, in good agreement with that derived from the present measurement.

In summary, we have performed the first investigation of the single-particle structure of the bound states of $^{17}$C using the $(d,p)$ reaction in inverse kinematics with a secondary beam of $^{16}$C.   These measurements have confirmed the spin-parities for the excited states of $1/2^{+}$ ($E_x=217$~keV) and $5/2^+$ ($E_x=335$~keV). Most significantly, both levels were found to carry very large single-particle strengths, whilst the spectroscopic factor for the $3/2^{+}$ ground state was deduced to be very small.  The $2s_{1/2}$ and $1d_{5/2}$ neutron single-particle orbits were thus shown to be essentially degenerate in $^{16,17}$C and consequently no $N=14$ sub-shell closure occurs.  In addition, the result for the ground state indicates that essentially all of the neutron $1d_{3/2}$ strength must lie in unbound states in $^{17}$C. 
The energies and spectroscopic factors obtained here were found to be in good agreement with shell model calculations using a range of interactions, including the recently developed SFO-tls and YSOX interactions, as well as a microscopic cluster model. Finally, the large $s$-wave spectroscopic strength for the $1/2^{+}$ level combined with its weak binding suggest that it is a good candidate for a neutron halo state.

\section*{Acknowledgments}
X.P.L. wishes to acknowledge the financial support of an IN2P3/CNRS (France) doctoral fellowship and the ST/P003885 grant (Spain). B.F.D. and M.C.F acknowledge financial support from the Ram\'on y Cajal programme RYC-2010-06484 and RYC-2012-11585 and from the Spanish MINECO grant No. FPA2013-46236-P. This work is partly supported by MINECO (Spain) grant 2011-AIC-D-2011-0802 and by the Xunta de Galicia through the grant EM2013/039. W.N.C. and A.M. acknowledge financial support from the STFC grant number ST/L005743/1. A. Moro and J.A. Lay acknowledge the Spanish Ministerio de Ciencia, Innovaci\'on y Universidades and FEDER funds under project FIS2017-88410-P and RTI2018-098117-B-C21 and the European Union's Horizon 2020 research and innovation program under Grant Agreement No. 654002. The authors acknowledge the support provided by the technical staff of LPC-Caen and GANIL. The participants from the Universities of Birmingham and Surrey, as well as the INFN and IFIN-HH laboratories also acknowledge partial support from the European Community within the FP6 contract EURONS RII3-CT-2004-06065.



\end{document}